\documentclass{jfm}
\usepackage{graphicx}
\usepackage[section]{placeins}
\usepackage{float}
\usepackage{epstopdf, epsfig}
\usepackage[utf8]{inputenc}
\usepackage{textcomp}
\usepackage{multirow}
\usepackage{graphicx}
\usepackage{epstopdf}
\usepackage{pdfpages}
\usepackage{soul}
\usepackage{xcolor}
\usepackage{natbib}
\usepackage{subfigure}
\usepackage{hyperref}
\usepackage{lineno}
\hypersetup{
	colorlinks = true,
	linkcolor  = blue,
	urlcolor   = blue,
	citecolor  = blue,
}

\shorttitle{finite-size particles in turbulent plane-Couette flow}
\shortauthor{C. Wang, L. Jiang and C. Sun}

\title{Numerical study on turbulence modulation of finite-size particles in plane-Couette flow}

\author{Cheng Wang\aff{1}, Linfeng Jiang\aff{1,2}\corresp{\email{l.jiang-1@utwente.nl}}, \and Chao Sun\aff{1,3}\corresp{\email{chaosun@tsinghua.edu.cn}}}

\affiliation{
	\aff{1}Center for Combustion Energy, Key Laboratory for Thermal Science and Power Engineering of Ministry of Education, International Joint Laboratory on Low Carbon Clean Energy Innovation, Department of Energy and Power Engineering, Tsinghua University, Beijing 100084, China
    \aff{2}Physics of Fluids Group, and Max Planck UT Center for Complex Fluid Dynamics, University of Twente, 7500 AE Enschede, The Netherlands
	\aff{3}Department of Engineering Mechanics, School of Aerospace Engineering, Tsinghua University, Beijing 100084, China
 }

\begin{document}

\maketitle

\begin{abstract}
Turbulent plane-Couette flow suspended with finite-size spheroidal particles is studied using fully particle-resolved direct numerical simulations.
The effects of particle aspect ratio on turbulent arguments and particle statistics are explored, leading to the same conclusions as the previous experimental findings \citep{wang2022finite}. 
By performing stress analysis, we find that the presence of particles introduces extra stresses to the system and accounts for the global drag increases. The particle-laden flow cases exhibit spectra that are consistent with the scaling of $k^{-5/3}$ and $k^{-3}$ in the large and small scales, respectively. 
While the $k^{-3}$ scaling observed in the particle-laden flow is reminiscent of bubbly flow, an examination of the particle Reynolds number suggests that the mechanism responsible may not be attributable to the pseudo-turbulence induced by particles as in the case of bubbles.
In the view of particle statistics, we observe that spherical and non-spherical particles preferentially cluster in the near-wall and the bulk region, respectively, and that the orientations of non-spherical particles are affected by their aspect ratios, especially in the near-wall region. The present numerical results, combined with previous experimental findings in \cite{wang2022finite}, provide in-depth information on both the fluid and the particle phase, contributing to a better understanding of particle suspension in shear flows.

\end{abstract}

\section{Introduction}
The intriguing landscape of particles dispersed in turbulent flows has attracted tremendous attention over the past decades due to their extensive existence in nature and industry \citep{pedley1992hydrodynamic,moffet2009situ,lundell2011fluid,mittal2020flow}. The presence of particles could introduce an extra scale (the particle size) to the turbulent flows, which may alter the energy cascade of the turbulence and modulate the turbulent flow. 
For neutrally buoyant particles as considered in the present work, when the suspension is dilute and/or the particles are extremely small \citep{voth2017anisotropic,mathai2020bubbly, brandt2022particle}, the particle-laden system could be simplified using the point-particle model by taking the dispersed particles as inertialess particles which could faithfully follow the surrounding flows. 
Such typical situations applying the point-particle model include plankton in the ocean \citep{stocker2012marine, qiu2022active} and pollen species in the atmosphere \citep{sabban2011measurements}. 
For particles with higher volume fractions or non-negligible inertia, the interactions between particle and fluid can be modelled as two-way coupling using point-particle method toward the motivation of investigating the particle dynamics and flow modulations (see the literature in the reviews \citep{voth2017anisotropic, brandt2022particle}).
However, given its simplicity, the point-particle model could not precisely account for the feedback of particles on the surrounding flows in general situations, such as the particle boundary layer and the wake behind them \citep{jiang2022dynamics}.
To this end, the finite-size particles, typically with diameters larger than the dissipation length scale $\eta$ of the surrounding turbulent flows 
\citep{voth2017anisotropic,mathai2020bubbly,brandt2022particle}, have been a vigorous field in recent years, both in experiments \citep{qureshi2007turbulent, qureshi2008acceleration, fiabane2012clustering, will2021kinematics, will2021dynamics, will2021rising, obligado2022dynamics} and numerical simulations \citep{calzavarini2009acceleration, picano2015turbulent, yousefi2020modulation,peng2020flow, demou2022turbulent, li2022inertial, assen2022strong}. 
In particular, by numerically implementing no-slip boundary conditions on their surface, the finite-size particles are capable of modelling the particle dynamics \citep{jiang2020rotation,jiang2022dynamics} and the resulting turbulence modulation \citep{wang2017fully,ardekani2017drag,ardekani2019turbulence}.

One of the subjects of particle-laden turbulence in the non-dilute regime is the turbulence modulation caused by particles. 
Due to the no-slip boundary conditions at their surface, the finite-size particles could modulate the surrounding flow and the entire field such as through shedding vortex in the wake region \citep{risso2018agitation,mathai2020bubbly} and enhancing the dissipation rate around them \citep{jiang2022dynamics}. 
For particle-laden turbulence with neutrally buoyant particles, the system can be characterized by four dimensionless numbers, namely, the flow Reynolds number $Re$, 
the size-ratio of particle diameter to turbulence dissipation length scale $d_v/\eta$, the aspect ratio of particles $\lambda$ and the volume-fraction of particles $\phi$.
Among others, the aspect ratio of particles has been found to affect the clustering effects and particle dynamics, and thus altering the magnitude of turbulence modulation. 
For instance, \cite{ardekani2019turbulence} found that the spherical particles could result in an overall drag enhancement, while the non-spherical particles might result in drag enhancement or reduction depending on their aspect ratios. 
Similar findings on the turbulence modulation caused by non-spherical are also reported in their previous work \citep{ardekani2017drag}. 
In the view of turbulence, the aspect ratios of particles are found to affect their effects on the turbulence fluctuation intensity, thus altering the stress contributions \citep{ardekani2019turbulence}.

In our previous experimental work in turbulent Taylor-Couette (TC) flow \citep{wang2022finite}, it has been observed that the suspended spherical particles could result in a larger drag increase than the non-spherical particles. 
By proposing a qualitative analysis of the stress balance, the drag increases caused by the suspended particles are explained. 
On the other hand, it is observed that the clustering effects of particles are affected by the aspect ratio of particles. The spherical particles show clustering near the walls, whereas the non-spherical particles preferentially cluster in the bulk region. 
Based on the previous numerical studies of finite-size particles \citep{ardekani2017drag, ardekani2019turbulence}, it is conjectured that the preferential clustering of particles could be responsible for their different magnitudes of turbulence modulation. 
However, limited by the experimental techniques, an in-depth and more quantitative analysis is absent then on several aspects. 
Specifically, (\romannumeral1) how and to what extent is the basic turbulent flow modulated by the suspended particles with different aspect ratios? (\romannumeral2) how are the particle statistics affected by their aspect ratios? (\romannumeral3) what is the difference in the turbulence modulation between the near-wall clustering of spherical particles and the bulk clustering of non-spherical particles?

We should note that, although similar phenomena have been reported for channel flows, the difference of shear flow from the channel flow could give rise to new physics and observations. For example, the findings on drag modulation in \cite{wang2022finite} indicate that, regardless of the particle aspect ratio, only drag increments is possible for particle-laden Taylor Coutte flow. However, numerical studies in channel flows have found drag reduction in the cases of oblate particles (e.g.\cite{ardekani2019turbulence}), leaving open the question of whether this is a physical feature of channel flow or simply a numerical artefact. 
To resolve this issue, we can perform simulations in similar shear flows to those used in experiments on Taylor-Couette flow. In fact, unlike the pressure-driven channel flow, which has a parabolic velocity profile and yields the maximum velocity and zero velocity gradient at the channel centre, the Couette flow shows a non-zero gradient at the domain centre. The difference in velocity profile could affect the velocity fluctuation and thereby the stress balance (drag) in the fluid phase. Additionally, the particle statistics are affected by the velocity profile of the flow as their dynamics are related to the velocity gradient tensor of the flow \citep{voth2017anisotropic}. Therefore, investigating the shear flows laden with particles is of great interest.

To this end, this work aims to conduct fully particle-resolved direct numerical simulations (PR-DNS) to provide a quantitative analysis of our previous experimental results and answer the aforementioned questions. 
Performing simulation in Taylor-Couette flow is computationally expensive because it requires not only processing cylindrical geometries but non-uniform gird spacing to resolve the boundary layers in the radial direction. Hence, we employ a configuration of plane-Couette flow to carry out our investigation, which is of Cartesian coordinates and can be efficiently implemented with a scheme based on the Lattice Boltzmann Method.
Using immersed boundary method (IBM) \citep{peskin2002immersed}, we can resolve the motion of each particle and its surrounding flow field, allowing us to uncover the physics of both particle dynamics and the resulting turbulence modulation. The rest of the paper is organized as follows: Section 2 describes the flow configuration and numerical schemes, in Section 3 we discuss the results, including the turbulent arguments and the particle statistics, and Section 4 gives final remarks on this work.

\section{Numerical Methodology}
\subsection{Configurations of flow and particles}\label{sec:configuration}

\begin{figure*}
	\centering
	\includegraphics[width=0.8\linewidth]{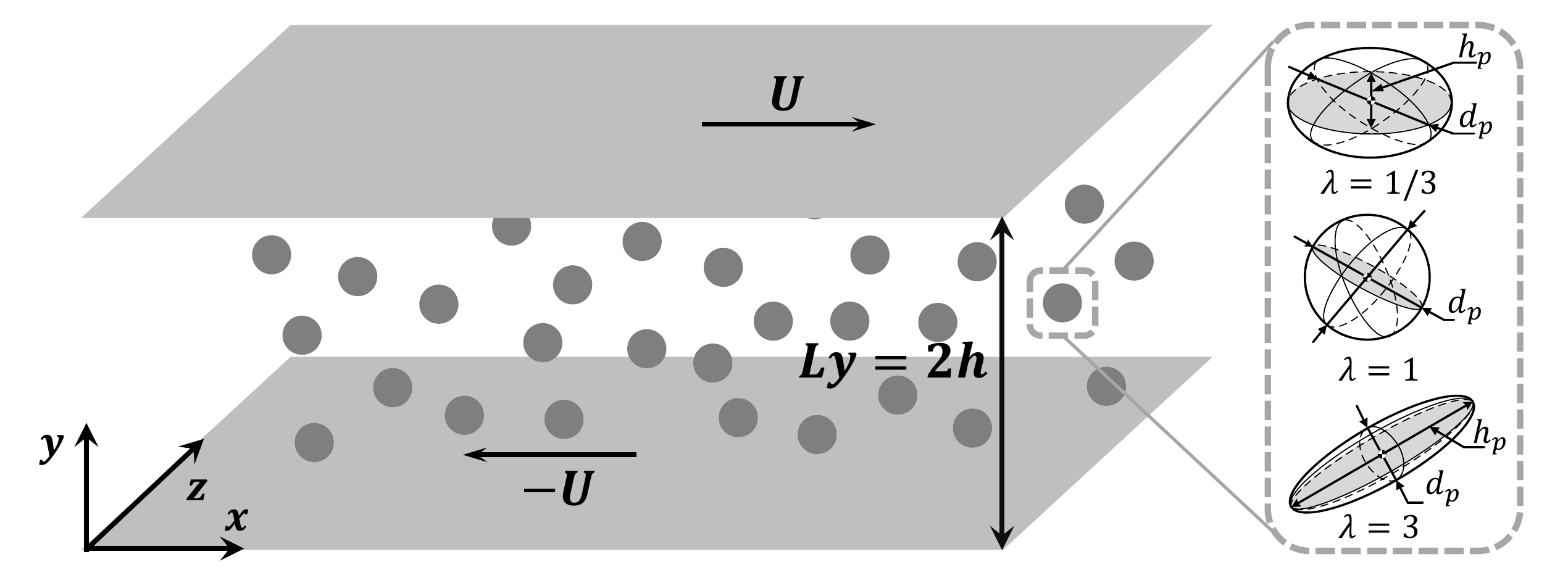}
	\vspace{2 mm}
	\caption{Sketch of the flow configuration. Two walls are moving at constant velocities in opposite directions. For particle-laden cases, particles with three kinds of aspect ratios are randomly suspended at the beginning and then move freely in the domain box.} 
	\label{fig1} 
\end{figure*}

We conduct simulations with spheroidal particles suspended in a turbulent plane-Couette flow. The flow configurations are depicted in figure~\ref{fig1}. Two walls are moving in the opposite direction with a constant velocity $U$. No-slip boundary conditions are imposed on the walls, while the flow in the streamwise and spanwise directions is periodic.
The turbulent flow is governed by the conservation equations of momentum and mass, which read as 
\begin{equation}\label{eq1}
    \partial_t \boldsymbol{u} + \boldsymbol{u}\cdot\nabla\boldsymbol{u} = -\rho^{-1}\nabla p +\nu \nabla^2\boldsymbol{u}+\boldsymbol{f}_p,
\end{equation}
\begin{equation}\label{eq2}
    \nabla \cdot \boldsymbol{u} = 0,
\end{equation}
where $\boldsymbol{u}$ and $p$ are the velocity vector and hydrodynamic pressure field of the flow, $\rho$ and $\nu$ are the density and viscosity of the fluid. Since the suspensions are composed of finite-size particles and in the non-dilute regime, the particle can modulate the surrounding turbulence by exerting feedback force $\boldsymbol{f}_p$ on the flow.
The turbulence intensity can be quantified by the bulk Reynolds number (hereafter, the Reynolds number), 
\begin{equation}\label{Re}
	Re = Uh/\nu,
\end{equation}
where $\nu$ is the kinematic viscosity of the fluid and $h$ is the half-width of the domain in the wall-normal direction.
Alternatively, the turbulence intensity can also be related to the shear Reynolds number, 
\begin{equation}\label{Ret}
	Re_\tau = u_\tau h/\nu,
\end{equation}
where $u_\tau = \sqrt{\tau_w/\rho}$ is the friction velocity with $\tau_w$ being the shear stress at walls. The values of $Re$ and the corresponding $Re_\tau$ are presented in Table~\ref{table1}. In this work, the values of $Re$ are set based on that in our previous experimental work. We note that in TC flow \citep{wang2022finite} the Reynolds number is defined as $Re_e = (u_i-u_o)d/\nu$, where $u_i$ ($u_o$) is the velocity of the inner (outer) wall and $d$ is the gap width between walls. Hence, the values of $Re$ in the present work are one-fourth of the corresponding values of $Re_e$ reported in \cite{wang2022finite} (see in Table~\ref{table1}).

\begin{table}
\centering
\renewcommand\tabcolsep{5.0pt}
\renewcommand{\arraystretch}{1.5}
\resizebox{1\linewidth}{!}{
\begin{tabular}{ccccccccccccccc}
$Re (=Re_e/4)$        & $Re_e$                  & $Nx \times Ny \times Nz$                     & $d_v/Ly$              & $\lambda$  &$\phi$ & $Re_\tau$  & $d_v/\eta$  & $Re^{slip}_p$ & $Re_p^{\lambda}$ & $\Delta/\delta_\nu$ & $\Delta/\eta$  & $u_\tau T/h$  & $\delta_{\tau_w}/\overline{\tau}_w$ \\
\multirow{7}{*}{1600} & \multirow{7}{*}{6400}   & \multirow{7}{*}{$384 \times 192 \times 192$} & \multirow{7}{*}{1/10} & -          & 0\%   & 99.8       &   -         & -             & -                & 1.04                & 0.658          & 78.0          & 0.7\%                               \\
                      &                         &                                              &                       & 1/3        & 2\%   & 99.3       & 12.64       & 11.06         & 64.0             & 1.03                & 0.658          & 76.9          & 1.5\%                               \\
                      &                         &                                              &                       & 1          & 2\%   & 101.3      & 12.74       & 6.38          & 64.0             & 1.06                & 0.664          & 145.1         & 0.4\%                               \\
                      &                         &                                              &                       & 3          & 2\%   & 98.9       & 12.58       & 1.92          & 64.0             & 1.03                & 0.655          & 83.0          & 0.5\%                               \\
                      &                         &                                              &                       & 1/3        & 6\%   & 100.4      & 13.00       & 2.15          & 64.0             & 1.05                & 0.677          & 80.4          & 3.0\%                               \\
                      &                         &                                              &                       & 1          & 6\%   & 105.2      & 13.02       & 6.36          & 64.0             & 1.10                & 0.679          & 89.7          & 0.4\%                               \\
                      &                         &                                              &                       & 3          & 6\%   & 100.5      & 12.78       & 4.94          & 64.0             & 1.05                & 0.666          & 81.8          & 0.4\%                               \\
\hline
\multirow{4}{*}{3200} & \multirow{4}{*}{12800}  & \multirow{4}{*}{$512 \times 256 \times 256$} & \multirow{4}{*}{1/10} & -          & 0\%   & 179.2      &   -         & -             & -                & 1.40                & 0.785          & 56.9          & 0.3\%                               \\
                      &                         &                                              &                       & 1/3        & 2\%   & 179.7      & 20.26       & 12.39         & 85.3             & 1.40                & 0.791          & 49.1          & 0.3\%                               \\
                      &                         &                                              &                       & 1          & 2\%   & 183.5      & 20.44       & 3.76          & 85.3             & 1.43                & 0.798          & 89.6          & 0.4\%                               \\
                      &                         &                                              &                       & 3          & 2\%   & 181.1      & 20.29       & 7.80          & 85.3             & 1.42                & 0.792          & 50.0          & 0.4\%                               \\
\end{tabular}
}
\caption{Simulation parameters. $Re$ is the bulk Reynolds number defined by the wall velocity $U$ and half-width of the gap $h$, which is therefore one-fourth of the corresponding values of $Re_e$ reported in \cite{wang2022finite} defined by the difference of wall velocity, $(u_i-u_o) = 2U$, and the gap width, $d=2h$. $Nx \times Ny \times Nz$ is the number of grids in each direction. $d_v/Ly$ and $d_v/\eta$ are the ratios of the equivalent-volume diameter of particles to the width of the domain in the wall-normal direction and to the dissipation length scale, respectively. $\lambda$ and $\phi$ are the aspect ratio and the volume fraction of particles. $Re_\tau$ is the friction Reynolds number defined by equation~\ref{Ret}. $Re^{slip}_p$ is the particle Reynolds number based on the global mean slip velocity and $Re_p^{\lambda}=Ud_v^2/h\nu^2$ is the particle Reynolds number based on local shear. $\Delta/\delta_\nu$ and $\Delta/\eta$ are the grid spacing normalized by the viscous length scale $\delta_\nu$ and the dissipation length scale $\eta$. $u_\tau T/h$ indicates the running times in terms of turnover periods for eddies of size $h$ and velocity $u_\tau$. $\delta_{\tau_w}/\overline{\tau}_w$ gives the statistic errors for each case, where $\delta_{\tau_w} = \tau_w|_{y=2h}-\tau_w|_{y=o}$ is the difference of wall stress, and $\overline{\tau}_w = (\tau_w|_{y=2h}+\tau_w|_{y=o})/2$ is the mean wall stress.}
\label{table1}
\end{table}

Neutrally buoyant spheroids are studied with different parameters, including the equivalent-volume diameter $d_v$, the aspect ratio $\lambda$, and the volume fraction $\phi$, chosen based on our previous experimental work \citep{wang2022finite}. 
Here for a spheroidal particle, we have $d_v = (h_pd_p^2)^{1/3}$ and $\lambda = h_p/d_p$, where $d_p$ is the length of the symmetric axis of the particle, and $h_p$ is the length perpendicular to it (see figure~\ref{fig1}).
The parameters of particles are given in Table~\ref{table1}. 
The particle motion is governed by the Newton-Euler equations as follows
\begin{equation}\label{NE1}
    m_p\frac{\mathrm{d} \boldsymbol{v}}{\mathrm{d}t} = \boldsymbol{F}+ \boldsymbol{F}_c,
\end{equation}
\begin{equation}\label{NE2}
    \frac{\mathrm{d} \boldsymbol{I}_p\boldsymbol{\Omega}}{\mathrm{d}t} = \boldsymbol{T}+ \boldsymbol{T}_c,
\end{equation}
where $\boldsymbol{v}(t) = \mathrm{d} \boldsymbol{r}/ \mathrm{d} t$ and $\boldsymbol{\Omega}(t)$ are the particle velocity and angular velocity vectors of a particle at position $\boldsymbol{r}(t)$ with mass $m_p = \rho_p V_p$ ($\rho_p$ is the particle density and $V_p$ the volume) and $\boldsymbol{I}_p$ the moment of inertia tensor. 
In the right-hand side (r.h.s.) of the above equations, $\boldsymbol{F}$ is the force exerted on particles by the surrounding fluid through hydrodynamic, 
while $\boldsymbol{F}_c$ is the collisions force accounting for the particles-particle collisions and particles-wall collisions. $\boldsymbol{T}$ and $\boldsymbol{T}_c$ are the torque defined in similar means. 

At $Re=1600$, two values of $\phi$ are chosen to study its effects on turbulence modulation. Whereas at $Re=3200$, only the case of small $\phi$ ($\phi = 2\%$) is conducted due to the costly computational expense at the high volume-fraction case, which nonetheless could uncover the roles of turbulent intensity.
For each $Re$ case, we first run the cases of single-phase flow (i.e., $\phi=0\%$) to generate a fully developed plane-Couette turbulence. 
Then the particle-laden case is conducted by randomly dispersing the particle in the entire domain with an initial velocity equal to zero. 
Statistical equilibrium is ensured by reaching time-averaged statistical errors in wall stress typically less than $1\%$ (see Table~\ref{table1}).

\subsection{Numerical Schemes}\label{sec:schemes}
In this section, we briefly describe the numerical schemes used to simulate the particle suspension. 
The turbulence is numerically solved using an open-source code with the lattice Boltzmann method (LBM), the $ch4-project$ \citep{calzavarini2019eulerian}, which has been extensively validated in particle-laden turbulence both for point-like particles \citep{mathai2016microbubbles,calzavarini2020anisotropic,jiang2021rotational} and finite-size particles \citep{jiang2020rotation,jiang2022dynamics}. 
The code was validated by comparing the translational dynamics of spheres at $Re_\lambda = 32$ with a previous reference study \citep{homann2010finite}. The details and  validations of the code can be found in our previous work \citep{jiang2022dynamics}.

On the fluid-phase side: The domain size is $Lx \times Ly \times Lz$ = $4h \times 2h \times 2h$, where $Lx, Ly, Lz$ are the width of the domain in the streamwise, wall-normal and spanwise direction, respectively. 
Here we note that the domain size is chosen based on producing fully-developed turbulence, which for the current $Re$ ranges is $Lx/Ly \simeq 2$ \citep{owolabi2018marginally}. To double-check if the domain size is long enough and the artefacts of periodic boundary conditions affect the results, we run an addition case with $Lx \times Ly \times Lz$ = $8h \times 2h \times 2h$. It is found that the drag coefficient ( see the $blue$ square dot in figure~\ref{fig2}a) and the particle statistics (not shown here) are comparable to the cases reported in Table~\ref{table1}. Therefore, the domain size in Table~\ref{table1} is long enough to yield conclusive results.
The computational domain is meshed by grids distributed uniformly in each direction, and the numbers of grids, $Nx \times Ny \times Nz$, are given in Table~\ref{table1}. 
Considering the requirement of the numerical resolution, the maximum grid spacing is validated to be $\Delta \leq 1.1\delta_\nu$ and $\Delta \leq 0.68\eta$ at $Re = 1600$ (see Table~\ref{table1}), where $\delta_\nu = \nu/u_\tau$ is the viscous length scale and $\eta = (\nu^3/\epsilon)^{1/4}$ is the dissipation length scale with $\epsilon$ being the global dissipation rate of the flow. 
In consequence, about 5 grids are embedded in the linear viscous layer (as can be seen in figure~\ref{fig5}). Hence, the grid spacing adopted in all runs is sufficient to resolve the turbulence fluctuation and flow structures, both in the boundary layers and the bulk. 

On the particle-phase side: the particle-fluid interaction is fully resolved by employing the immersed boundary method (IBM) \citep{peskin2002immersed}.  
The IBM enforces the no-penetration and no-slip boundary conditions at the fluid–particle interface by means of a localized feedback force, $\boldsymbol{f}_p$, added to equation~\ref{eq1}. Such $\boldsymbol{f}_p$ term is also denoted as two-way coupling.
The particle surfaces are captured by uniformly distributed Lagrangian nodes. Adequate grids are required to resolve the flow around particles. The test toward this has been carried out in our previous work \citep{jiang2022dynamics}. It is found that at least 16 grids per particle diameter are needed to resolve the particle boundary layers. In this work, we surpass this criterion to attain convincing results. Specifically, for the cases of $Re=1600$ and $Re=3200$, 19 and 26 grids per particle diameter are respectively used to resolve the particle boundary layers and their dynamic.
To ensure high accuracy for the implementation of the no-slip fluid boundary condition at the particle surface, we adopt the so-called IBM multi-forcing scheme with 5-step iterations, see \cite{luo2007full} and \cite{wang2021heat}. 
The processes of particle-particle collisions and particle-wall collisions are implemented by means of soft-sphere collision forces \citep{costa2015collision,ardekani2016numerical} and lubrication corrections \citep{brenner1961slow, cooley1969slow, costa2015collision, ardekani2016numerical}, where $\boldsymbol{F}_c$ and $\boldsymbol{T}_c$ can be computed. 
The details of the calculations of the hydrodynamic force and torque can be found in our previous work \citep{jiang2022dynamics}.
In our simulations with the immersed boundary method, the boundary of the finite-size particle is represented by Lagrangian nodes and the particle-fluid interactions are solved by the delta function as suggested by Peskin. Furthermore, we seed Lagrangian points inside the finite-size particle which are fixed in the particle frame. These Lagrangian points are used to measure the momentum time derivative of fluid inside the finite-size particle to obtain the correct driving force. And we use the tri-linear scheme to interpolate the velocities of the Lagrangian points inside the finite-size particle.
The Newton-Euler equations that govern the particle dynamics are numerically integrated with a second-order Adams–Bashforth time-stepping scheme.

Since the size-ratio $d_p/\eta$ is fixed, the numbers of particles $Np$ are determined by their volume fractions, yielding $Np=76$ and $229$ for the cases of $\phi=2\%$ and $\phi=6\%$, respectively. Therefore, the statistical convergence of the numerical wall stress and the particle statistics are ensured by running the simulation long enough to get sufficient data. The running time for each case is given in Table~\ref{table1}. For additional information, the time series of the fluid and particle velocity fluctuations are shown in Appendix~\ref{sec:appendix}.

\section{Results and Discussion}
\subsection{Turbulence modulation: global drag, velocity field, and dissipation rate}\label{sec: first}
\begin{figure*}
	\centering
	\includegraphics[width=1\linewidth]{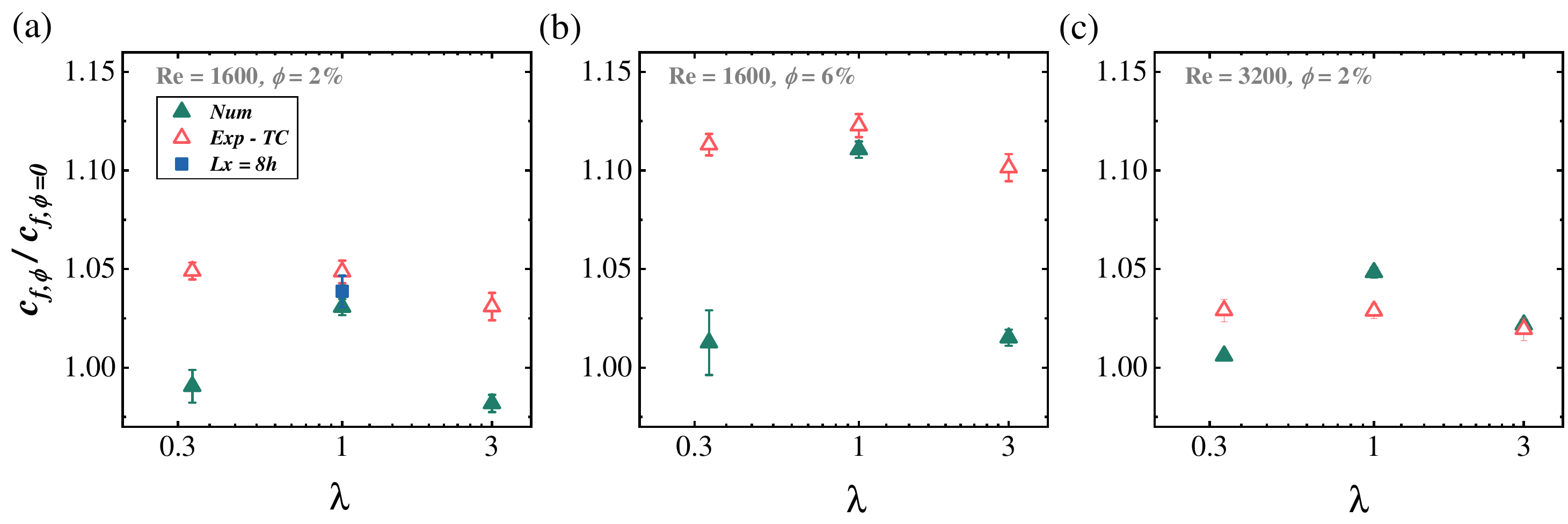}
	\vspace{-5 mm}
	\caption{Normalized wall stress for particle-laden cases. The experimental data (empty circles) is adapted from the TC experiments \citep{wang2022finite}. The $blue$ square in (a) denotes the result of the validation case of extended domain size in the streamwise direction ($Lx \times Ly \times Lz$ = $8h \times 2h \times 2h$). The values of $Re$ and $\phi$ are shown in each panel.}
	\label{fig2} 
\end{figure*}

We start by comparing the global turbulence modulation of numerical results to that of experimental measurements \citep{wang2022finite}. The global transport quantity of plane-Couette turbulence is the shear stress at the walls 
\begin{equation}\label{tauw}
    \tau_{w,\phi} = \rho\nu \frac{\mathrm{d}U_f}{\mathrm{d}y}|_{y=0,2h},
\end{equation}
where $\frac{\mathrm{d}U_f}{\mathrm{d}y}|_{y=0,2h}$ is the gradient of mean stream-wise velocity at the walls. 
We normalize the wall stress of particle-laden flow by that of single-phase flow, i.e., $\tau_{w,\phi}/\tau_{w,\phi=0}$, and show it in figure~\ref{fig2}, where the results from previous experiments in Taylor-Couette flow\citep{wang2022finite} are also depicted for comparison.
The discrepancies between numerics and experiments could be attributed to the different flow configurations.
Although both flow configurations are shear-induced flows, TC flow suffers from more complex conditions, such as the cylindrical geometries causing asymmetric velocity profiles in the radial direction, and the strong confinement due to the end-plates which cause secondary flow \citep{grossmann2016high}. It is therefore not expected to observe exact agreements between the results of numeric and experiments. 
Moreover, as the particle dynamics are affected by their aspect ratios \citep{voth2017anisotropic}, the spherical and non-spherical particles could respond in different means as the flow configuration changes, therefore resulting in different responses in drag modulation. This could possibly account for that in drag modulation (figure~\ref{fig2}), compared to non-spherical particles, the spherical particles show better agreements in TC flow and plane-Coutte simulations.
Nevertheless, concerning that we are mainly interested in the effects of particle aspect ratio on the turbulence modulation, the conclusion can be drawn that the spherical particle could cause larger drag modulations than the non-spherical particles, for different $Re$ and $\phi$ with the numerical results in figure~\ref{fig2}. Similar findings are also reported in our previous experimental TC flows, where the difference in magnitude of drag modulation between spherical and non-spherical cases is not obvious at the present $\phi$ ($\Delta_{c_f/c_{f,\phi=0}}\sim 1\%$, see figure 2) but become larger at a higher value of $\phi$ ($\Delta_{c_f/c_{f,\phi=0}}\sim 5\%$ at $\phi=10\%$).

\begin{figure*}
	\centering
	\includegraphics[width=1\linewidth]{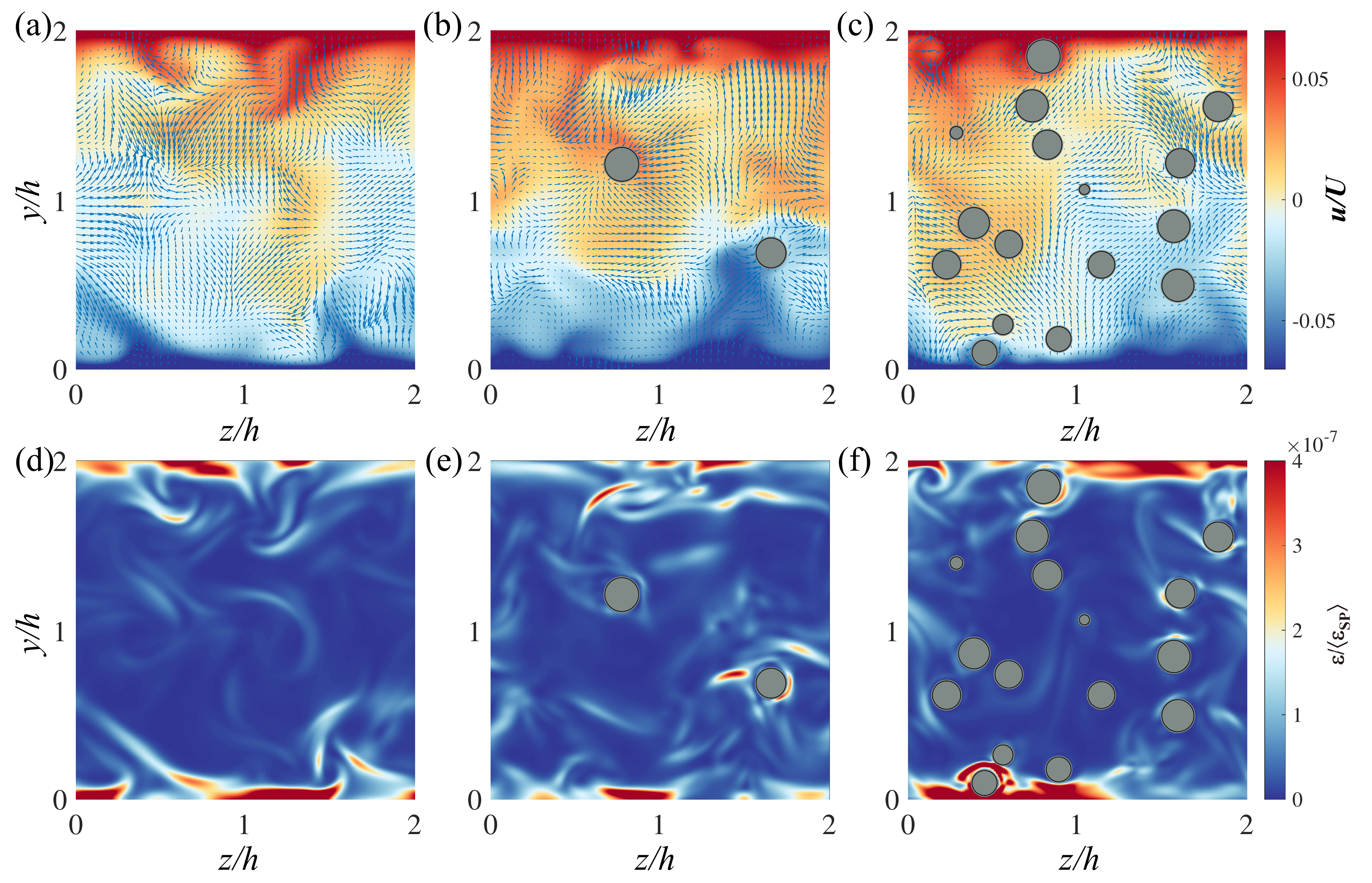}
	\vspace{-5 mm}
	\caption{Instantaneous snapshots of (a-c) stream-wise velocity indicated by the colour bar. The overlaid arrow vectors depict flow velocity within the plane. (d-f) dissipation rate. Data is obtained from the cases of spherical particles ($\lambda = 1$) at $Re=1600$. Columns from left to right are $\phi=0\%$, $\phi=2\%$ and $\phi=6\%$, respectively.} 
	\label{fig3} 
\end{figure*}

\begin{figure*}
	\centering
	\includegraphics[width=1\linewidth]{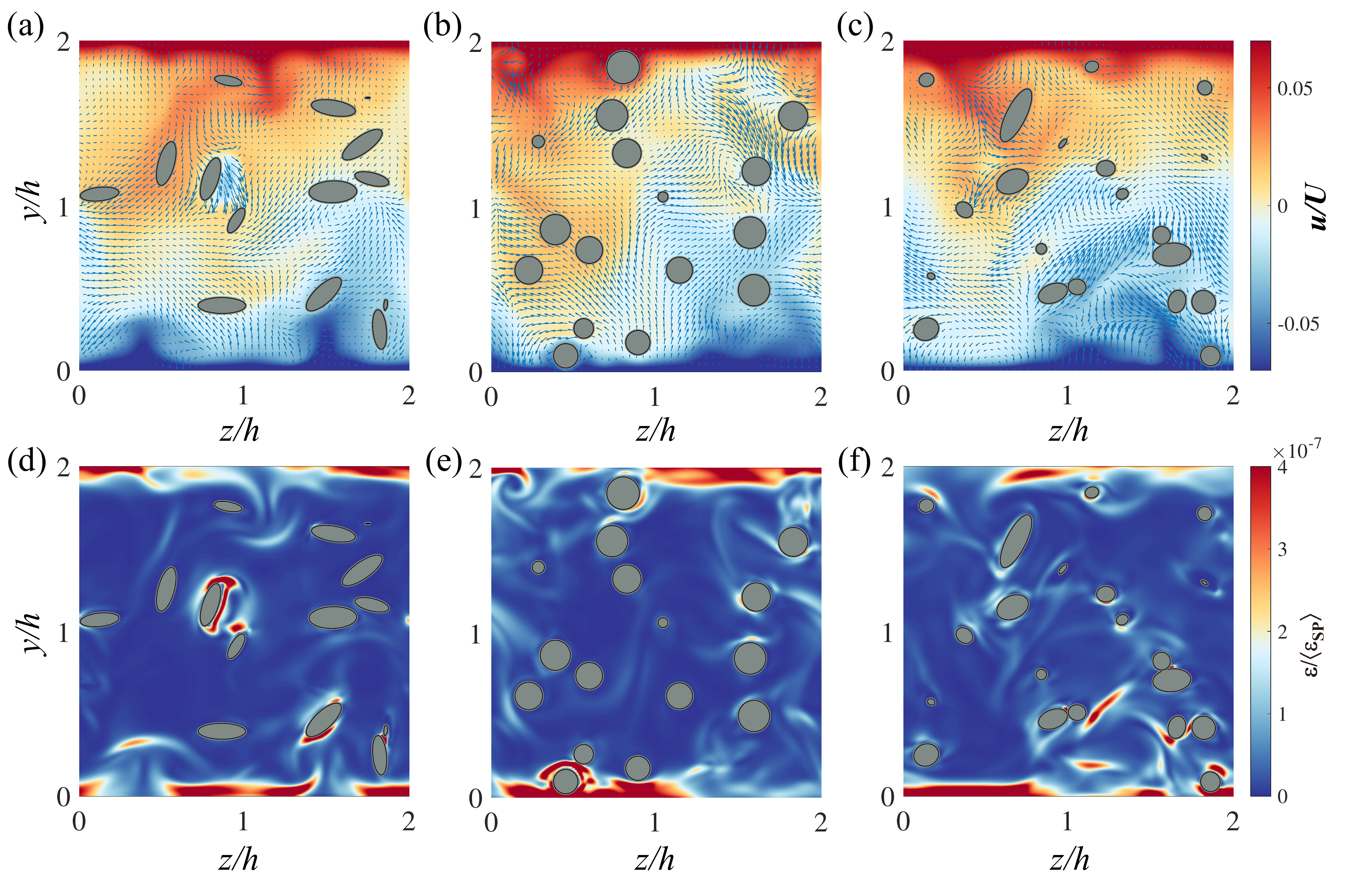}
	\vspace{-5 mm}
	\caption{Instantaneous snapshots of (a-c) stream-wise velocity indicated by the colour bar. The overlaid arrow vectors depict flow velocity within the plane. and (d-f) dissipation rate. Data is obtained at $Re=1600$ and $\phi=6\%$. Columns from left to right are for $\lambda = 1/3$, $\lambda = 1$ and $\lambda = 3$, respectively.} 
	\label{fig4} 
\end{figure*}

For fixed $Re$ and $\lambda$ (see figures~\ref{fig2}(a, b)), the magnitudes of drag modulation increase with increasing $\phi$, which again agrees with the results in our previous experimental measurements \citep{wang2022finite}. 
To demonstrate this $\phi$-dependence of drag modulation, in figure~\ref{fig3} we present the velocity field and the corresponding fields of dissipation rate at the centre plane in stream-wise direction (i.e. $x=Lx/2$). 
For clarity, only the case of the spherical particles at $Re=1600$ is shown. 
Compared to the case of single-phase (figure~\ref{fig3}a), the presence of particles could introduce strong disturbance both in the bulk and the boundary layers, see the emergence of the higher value of dissipation rate coloured in red around the particles in figures~\ref{fig3}(b,c).
On the other hand, figure~\ref{fig4} shows similar information as figure~\ref{fig3} but for different $\lambda$ at $Re=1600$ and $\phi=6\%$. It can be seen that the particle could modify the surrounding fields and particularly enhance the dissipation rate near their surface, which again could be related to the particle boundary layers \citep{jiang2022dynamics} and accounts for the drag increases reported in figure~\ref{fig2}. 
This enhancement of turbulence dissipation of particles has been found in other flows, e.g. homogenous and/or isotropic turbulence \citep{cisse2013slipping,de2016local,jiang2022dynamics}, and decaying isotropic turbulence \citep{lucci2010modulation}. Due to the no-slip condition enforced at the particle surface, a local force arises on the fluid surrounding the particle, which locally increases the velocity gradients close to the particle surface and thus increases the local strain rate as well as the dissipation rate \citep{lucci2010modulation}. As the distance to the particle surface increases, this impact of the no-slip condition fades, and so does the dissipation rate. This accounts for the observations in figures~\ref{fig3}, \ref{fig4}, where the bursts of high dissipation rate (coloured in $red$) only occur near the particles while the rear fluid almost remains undisturbed (coloured in $blue$). Indeed, the particles are found to enhance the dissipation rate around them to a distance of about one radius to their surfaces \citep{cisse2013slipping,de2016local} for spherical particles, while for non-spherical particles the volume of disturbed fluid is smaller in every dimension \citep{jiang2022dynamics}. Additionally, the enhancement of the dissipation rate around particles distributes in a non-uniform manner, namely, the dissipation rate in front of the particle is higher than the rear of the particles \citep{cisse2013slipping,de2016local}. Although this can not be quantified in the present work, one can still be hinted from figures~\ref{fig3}(e,f) and figures~\ref{fig4}(d-f), where the dissipation rate around particles is asymmetric. With increasing volume fraction of particles, the dissipation rate becomes pronounced, as could be seen in figures~\ref{fig3},\ref{fig4}. This is due to the increase of particle surface area that affects the surrounding fluid and thus the two-way coupling force \citep{lucci2010modulation}.
Moreover, a closer look would find that the spherical particles cause stronger disturbances to the boundary layers, while the non-spherical particles mainly alter the flow in the bulk. 

\begin{figure*}
	\centering
	\includegraphics[width=1\linewidth]{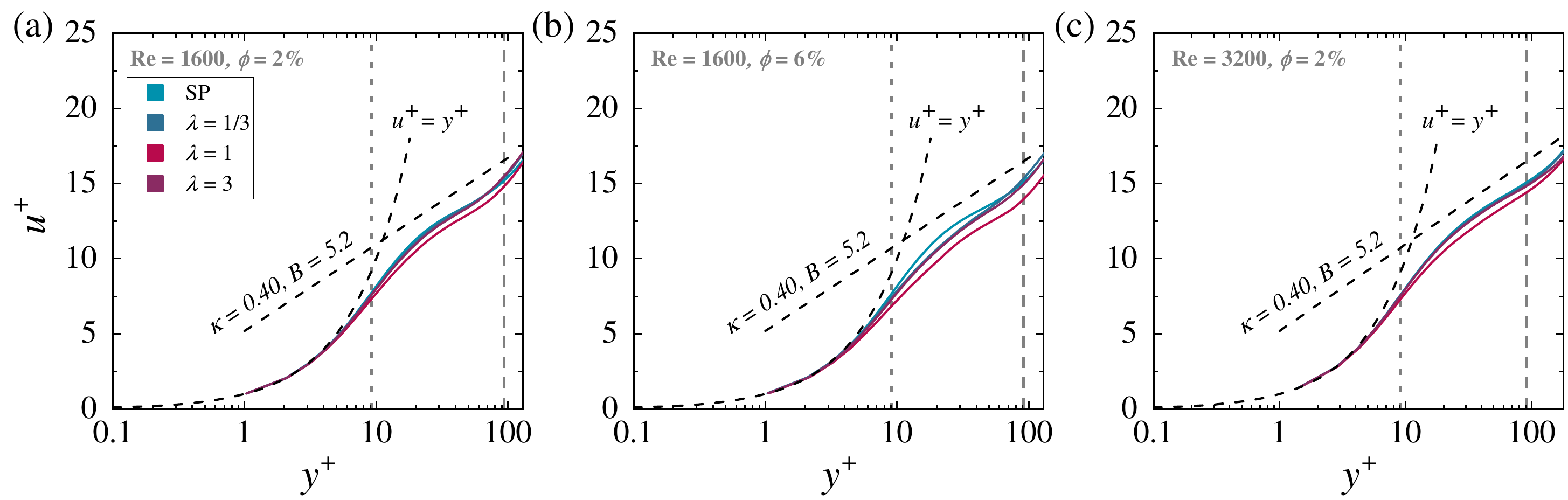}
	\vspace{-5 mm}
	\caption{Velocity profiles depicted in the wall units. The vertical dotted and dashed lines in each panel indicate, respectively, the location of the equivalent-volume radius of particles ($r_v=d_v/2$) and the location of the central plane in the wall-normal direction ($y=h$).}
	\label{fig5} 
\end{figure*}

\begin{figure*}
	\centering
	\includegraphics[width=1\linewidth]{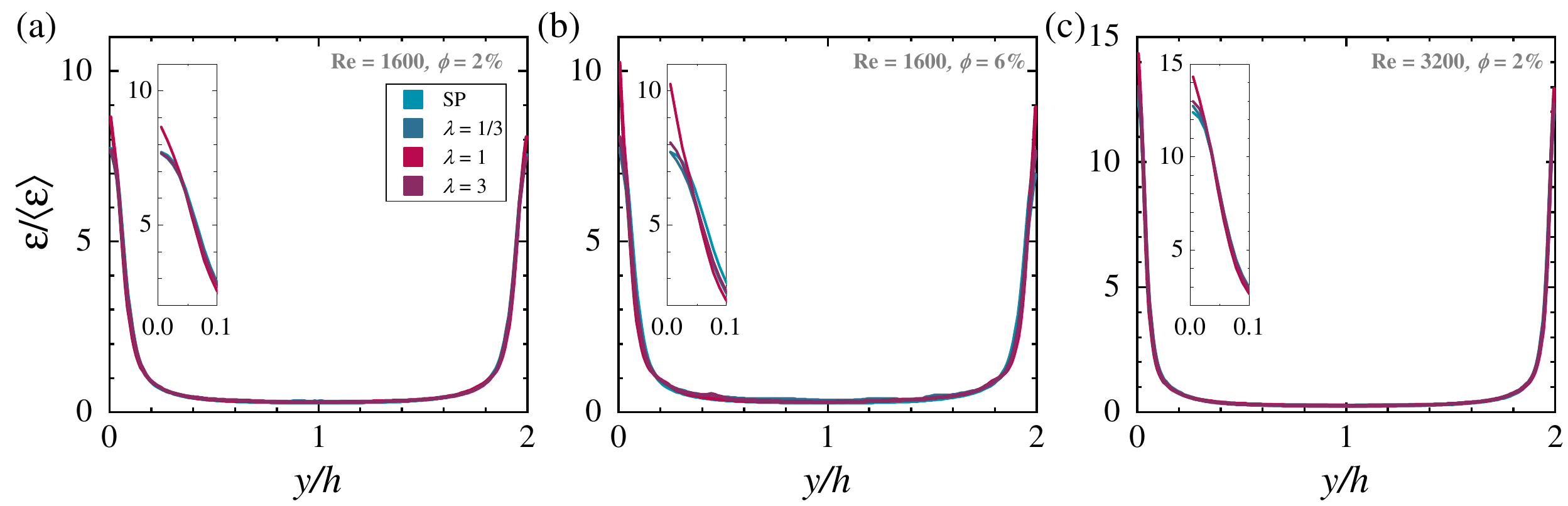}
	\vspace{-5 mm}
	\caption{Profiles of dissipation rate. The insets show the zoom-in view of the dissipation rate in the near-wall region.} 
	\label{fig6} 
\end{figure*}

We depict the profiles in the wall-normal direction of stream-wise velocity and dissipation rate in figure~\ref{fig5} and figure~\ref{fig6}, respectively. 
In figure~\ref{fig5}, the velocity profile is shown in the wall units, where $y^+ = y/\delta_\nu$ is the distance from the lower wall ($y=0$) in units of the viscous length scale $\delta_\nu$, and $u^+ = (u-u|_{y=0})/u_\tau$ is the velocity difference from the lower wall normalized by the friction velocity $u_\tau$. 
For comparison, we also depict the viscous sublayer $u^+=y^+$ and the logarithmic law of flat turbulent boundary layer flow, i.e. $u^+= 1/\kappa~${ln}$~y^++B$ with the typical values of $\kappa=0.40$ and $B=5.2$ as suggested by Prandtl and von-K\'arm\'an \citep{pope2000turbulent}. 
One can see that, regardless of the values of $\phi$ and $\lambda$, the profile of $u^+=y^+$ is well retained with sufficient grids inside the viscous sublayer ($y^+<5$), which is barely affected by the suspended particles due to their finite sizes (see the vertical dotted lines in figure~\ref{fig5}). 
Outside the viscous sublayer, the velocity profiles generally follow the logarithmic law in all cases. 
The effects of particle aspect ratio become pronounced with increasing $\phi$ (see figure~\ref{fig5}b). 
Compared to single-phase flow, the cases of particle-laden flow result in the velocity profile consisting of an extended logarithmic layer, which is similar to the observation in channel flow \citep{eshghinejadfard2018lattice} and yields in the reduced velocity gradient in the near-wall region.
This seems to contradict the drag increases reported in figure~\ref{fig2}. However, as shown later, for particle-laden flow the dampening in velocity does not imply a global drag reduction since there are extra stress contributions from the particle phase. 
Moreover, in terms of the effect of particle aspect ratio, it is found that at fixed $Re$ and $\phi$ the spherical particles cause greater velocity reduction than the non-spherical particles in the logarithmic layer (e.g., see figure~\ref{fig5}b). 
Specifically, in figure~\ref{fig5}b, the value of $B$ of spherical particles ($B\simeq 2.3$) is smaller than that of non-spherical particles ($B\simeq 3.5$) while the values of $\kappa$ are close to $0.40$ in all particle-laden cases. This is consistent with the drag increase found in figure~\ref{fig2} since the decrease in $B$ indicates a drag enhancement and the decrease in $\kappa$ does the opposite \citep{ardekani2017drag}.

As for the dissipation rate (figure~\ref{fig6}), its magnitudes agree well with the magnitudes of drag increases found in cases of different particle aspect ratios. We note that, though the non-spherical particles could alter the bulk flow stronger than the spherical ones in the bulk region (see figure~\ref{fig4}), the dissipation rates in the bulk are minor contributions to the total field, resulting in negligible changes of the profiles of dissipation rate in that region (figure~\ref{fig6}). In contrast, spherical particles efficiently increase the dissipation rate more than non-spherical particles near the walls (see insets in figure~\ref{fig6}). This is in line with the magnitudes of global drag modulation of particles with different $\lambda$ because, as the wall stress does, the dissipation rate in the near-wall region is positively associated with the velocity gradient at the walls.

\subsection{Turbulence modulation: stress analysis and turbulent energy spectrum}\label{sec:second}

In the previous work \citep{wang2022finite}, we decompose the total stress (i.e., the shear stress at walls $\tau_{w,\phi}$) into four terms, namely, the viscous stress of the fluid phase ($\tau_\nu$), the turbulent stress of fluid and particle phase ($\tau_{Tf}$ and $\tau_{Tp}$, respectively), and the particle-induced stress ($\tau_p$), which can be written as
\begin{equation}\label{stressbudget}
	\tau_{w,\phi} = \tau_\nu+\tau_{Tf}+\tau_{Tp} +\tau_{p}.
\end{equation}
where $\tau_{w,\phi}$ is defined by equation~\ref{tauw}. For each term in the r.h.s, one can explicitly express it, as suggested by \citep{batchelor1970stress, zhang2010physics, picano2015turbulent, wang2017modulation,wang2022finite}, in the form of turbulence quantities as,
\begin{equation}\label{taufv}
    \tau_{\nu} = \rho\nu (1-\phi) \frac{\mathrm{d} U_f}{\mathrm{d} y}
\end{equation}
\begin{equation}\label{tauft}
	\tau_{Tf} = -\rho(1-\phi) \langle u^{\prime}_fv^{\prime}_f \rangle,
\end{equation}
\begin{equation}\label{taupt}
	\tau_{Tp} = -\rho \phi \langle u^{\prime}_pv^{\prime}_p \rangle,
\end{equation}
\begin{equation}\label{taups}
	\tau_{p} = \phi \langle \sigma^p_{xy} \rangle,
\end{equation}
where $u^{\prime}$ and $v^{\prime}$ are the velocity fluctuation in the stream-wise and wall-normal direction with the subscripts $f$ and $p$ denoting fluid and particle phase, respectively, $\sigma^p_{xy}$ the general stress in the particle-phase, projected in the stream-wise plane and pointing in the stream-wise direction. $\langle ... \rangle$ means ensemble averaged in time and over the stream-wise plane. For the case of single-phase flow, the latter two terms are zero.

\begin{figure*}
	\centering
	\includegraphics[width=1\linewidth]{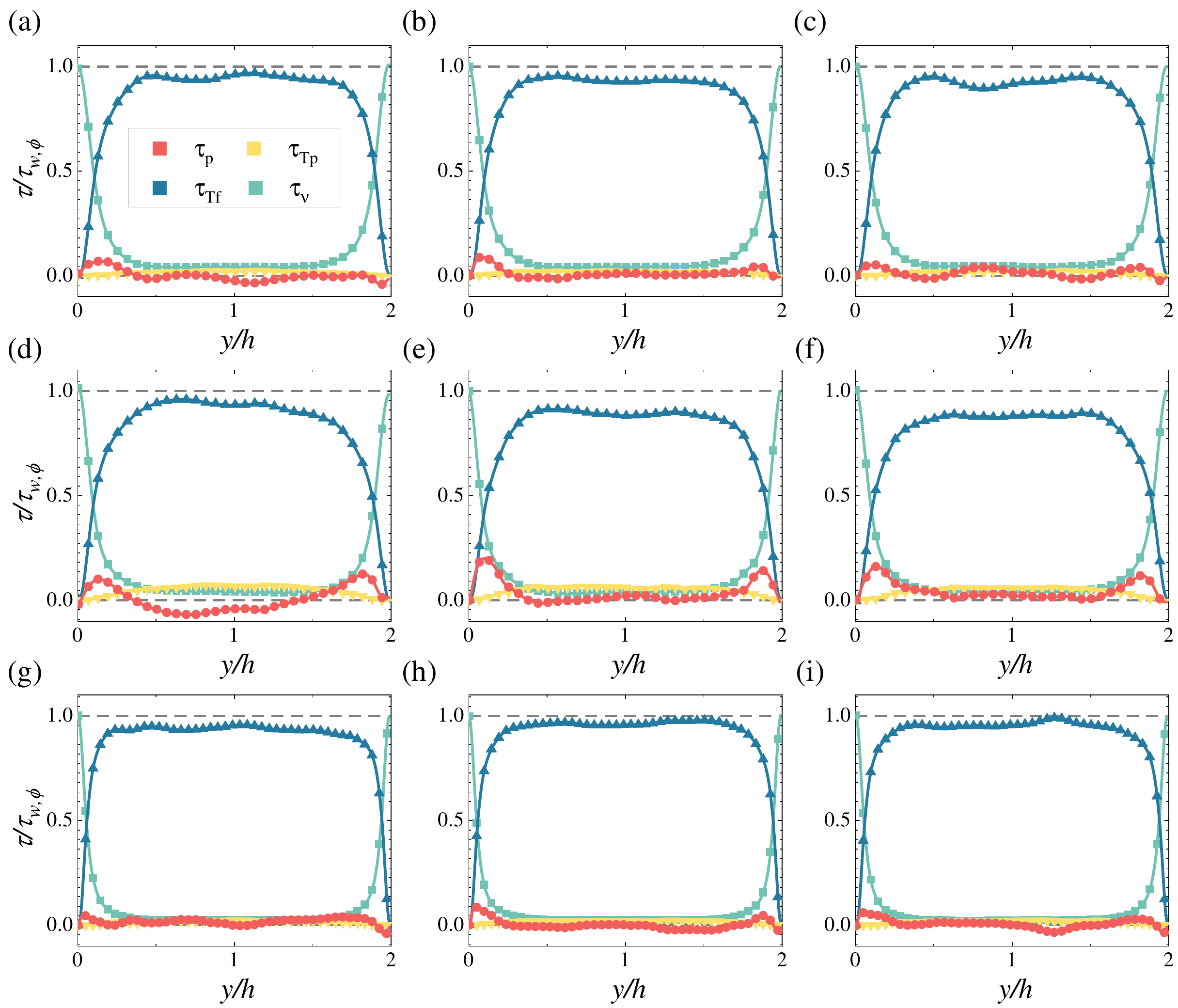}
	\vspace{-5 mm}
	\caption{Profiles of stress budget. Each term in r.h.s. of equation~\ref{stressbudget} is normalized by the total shearing stress of corresponding particle-laden cases (i.e. $\tau_{w,\phi}$ in equation~\ref{stressbudget}). The two horizontal dashed lines with the values of 1 and 0 are shown for reference.} 
	\label{fig7} 
\end{figure*}

By means of numerics, we can look into the contributions of each term in r.h.s of equation~\ref{stressbudget} to the total stress. 
The particle-induced stress $\tau_p$ is calculated indirectly (i.e. $\tau_p = \tau_w-\tau_\nu-\tau_{Tf}-\tau_{Tp}$) as the explicit expression of $\sigma^p_{xy}$ remains unknow currently.
In figure~\ref{fig7} we plot the profile of stress balance in the wall-normal direction for each case, where we normalize each term of stress by the total stress of that case, i.e., $\tau/\tau_{w,\phi}$. 
We first demonstrate how the particle aspect ratio, $\lambda$, affects the contributions of each stress at given $Re$ and $\phi$ (see each row of figure~\ref{fig7}). 
Taking the case of $Re=1600$ and $\phi=2\%$ as an example (first row of figures~\ref{fig7}), one can see that the particle aspect ratio has slight effects on $\tau_\nu$ and $\tau_{Tf}$. 
Similarly, the turbulent stress of the particle phase, $\tau_{Tp}$, only weakly depends on the $\lambda$ while having a positive dependence on the $\phi$ (see the second row of figure~\ref{fig7}). 
However, the particle-induced stress, $\tau_p$, is affected by the particle aspect ratios to a noticeable extent, which becomes even more pronounced at high $\phi$ (see the second row of figure~\ref{fig7}). 
Since $\tau_p$ is the combined results of hydrodynamic interaction and particle collisions \citep{picano2015turbulent}, its value could be related to both the particle dynamics and the turbulence field.
As the numerical errors from $\tau_p$ cannot be excluded now, we then take the case of high $\phi$  (second row of figures~\ref{fig7}), where the numerical errors are of less significance, as an instance to illustrate the effects of $\lambda$ on $\tau_p$. 
First, it is noticeable that in all cases of $\lambda$, $\tau_p$ reaches its peak near the walls. Near the walls, the spherical particles cause more pronounced peaks than the non-spherical particles in $\tau_p$, which validates our conjectures in previous work \citep{wang2022finite} and is in line with the findings in other configurations of particle-laden flow \citep{picano2015turbulent,ardekani2017drag}. 
As shown by the particle statistics in the later section, for spherical cases the occurrence of near-wall peaking in $\tau_p$ could originate from their near-wall preferential clustering, which could cause the particle collisions to happen in easier and more intense ways. 

\begin{figure*}
	\centering
	\includegraphics[width=1\linewidth]{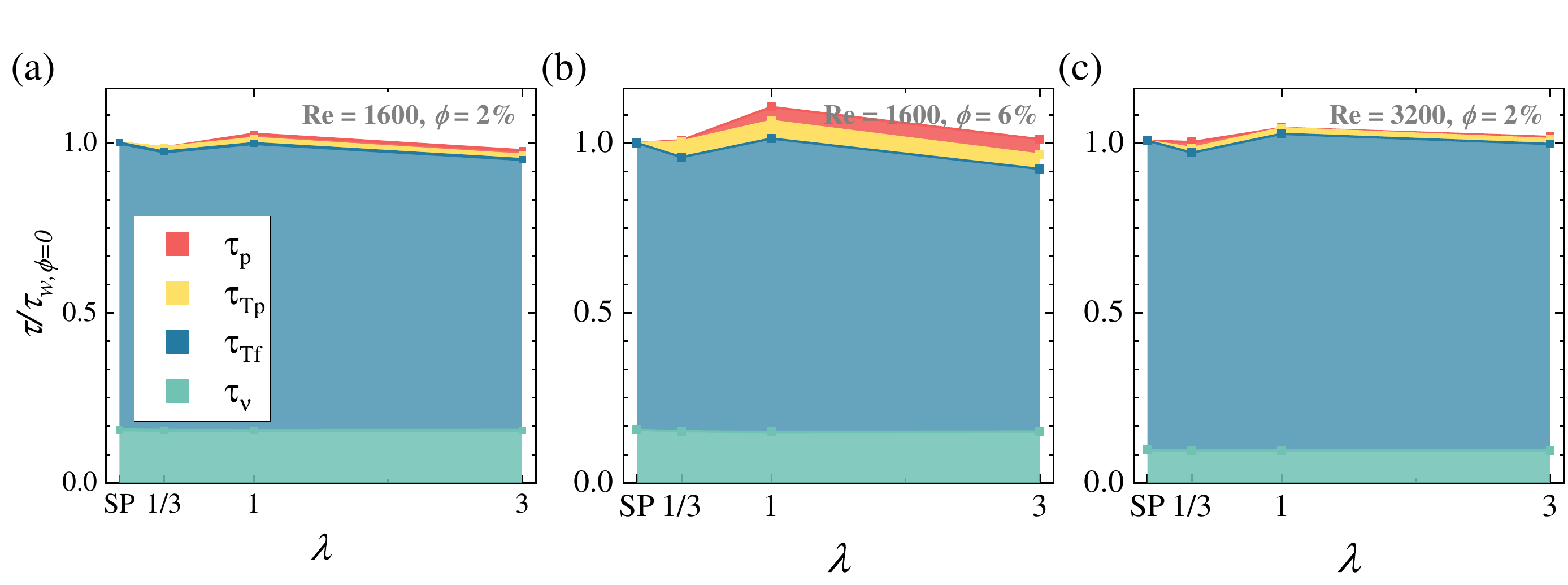}
	\vspace{-5 mm}
	\caption{Contributions of each individual stress to the total shearing stress. The total shearing stress of the corresponding cases of single-phase (i.e., $\tau_{w,\phi=0}$ at given $Re$) is used for normalization.} 
	\label{fig8} 
\end{figure*}

It is noticeable that, for the case of oblate ($\lambda=1/3$, see figure~\ref{fig7}d), $\tau_p$ becomes negative in the bulk region.
As the results are ensured to be convergence by running the simulation long enough in time, the fact that it is negative could be attributed to the hydrodynamic interaction caused by particles. This indicates that $\tau_p$ could be a stress sink under specific conditions. Similar overall drag reductions caused by particles have also been reported in simulation results in channel flow \citep{ardekani2019turbulence}. However, we should note that the previous experimental results \citep{wang2022finite} report that only drag enhancement is observed in particle-laden flow under similar parameter regimes, even for the cases of oblate particles. This is in contrast to the current simulation results, where the particle-induced stress is found to be negative (figure~\ref{fig7}d) and overall a drag (stress) reduction is observed (figure~\ref{fig8}). In this sense, the particle-induced stress being negative in the bulk could be a numerical artefact, though there are differences in TC and plance-Couette flows as we mentioned before. Future study is needed to uncover the role of $\tau_p$.
Moreover, the profiles in figure~\ref{fig7} are not exactly symmetric, especially in the particle-induced stress ($\tau_p$). This could be due to the limitations of statistics as the particle number determined by its volume fraction are relatively small, which could cause inhomogeneity in the bounded wall-normal direction (i.e. $y-$direction). Although extending the simulation time steps might improve the data quality, as we did for the spherical cases (see the values of $u_\tau T /h$ in Table 2), the asymmetries are still present in the spherical cases. Nonetheless, the asymmetric amplitude is typically less than 5\% for all simulation cases and thereby the main conclusion on the stress analysis can be convinced. 
On the other hand, as $Re$ increases (see the third row of figure~\ref{fig7}), the profiles of stress coming from the particle phase ($\tau_{Tp}$ and $\tau_p$) become almost flat. 
In these cases, the total stress is dominated by the viscous stress in the boundary layers and by the fluid-phase turbulence stress in the bulk, which is actually reminiscent of the case of single-phase flow.

The contribution ratio of each stress to the total one can be found in figure~\ref{fig8}, where we normalize each stress by the total shear stress at the walls of the cases of single-phase, i.e., $\tau/\tau_{w,\phi=0}$. 
For small $\phi$ (see figures~\ref{fig8}(a,c)), it can be seen that the total stress is almost dominated by the viscous stress and the turbulent stress, while the presence of particles negligibly affects the stress of the flow. 
The changes in $Re$ mainly redistribute the contributions from the $\tau_\nu$ and $\tau_{Tf}$ to the total stress. 
With increasing $\phi$ (see figure~\ref{fig8}b), the role of particles becomes pronounced as indicated by the greater contributions of $\tau_{Tp}$ and $\tau_p$. 
Indeed, one can see that the stress coming from the particle phase could principally be responsible for the drag increases found in the cases of particle-laden flow, even though the turbulent stress of the fluid phase might be dampened a bit by the suspended particles at higher $\phi$. 
Nevertheless, we note that, compared to the fluid phase, the contributions from particle-phase stress ($\tau_{Tp}$ and $\tau_p$) are noticeable but still minor effects due to the rather small $\phi$ here.

\begin{figure*}
	\centering
	\includegraphics[width=1\linewidth]{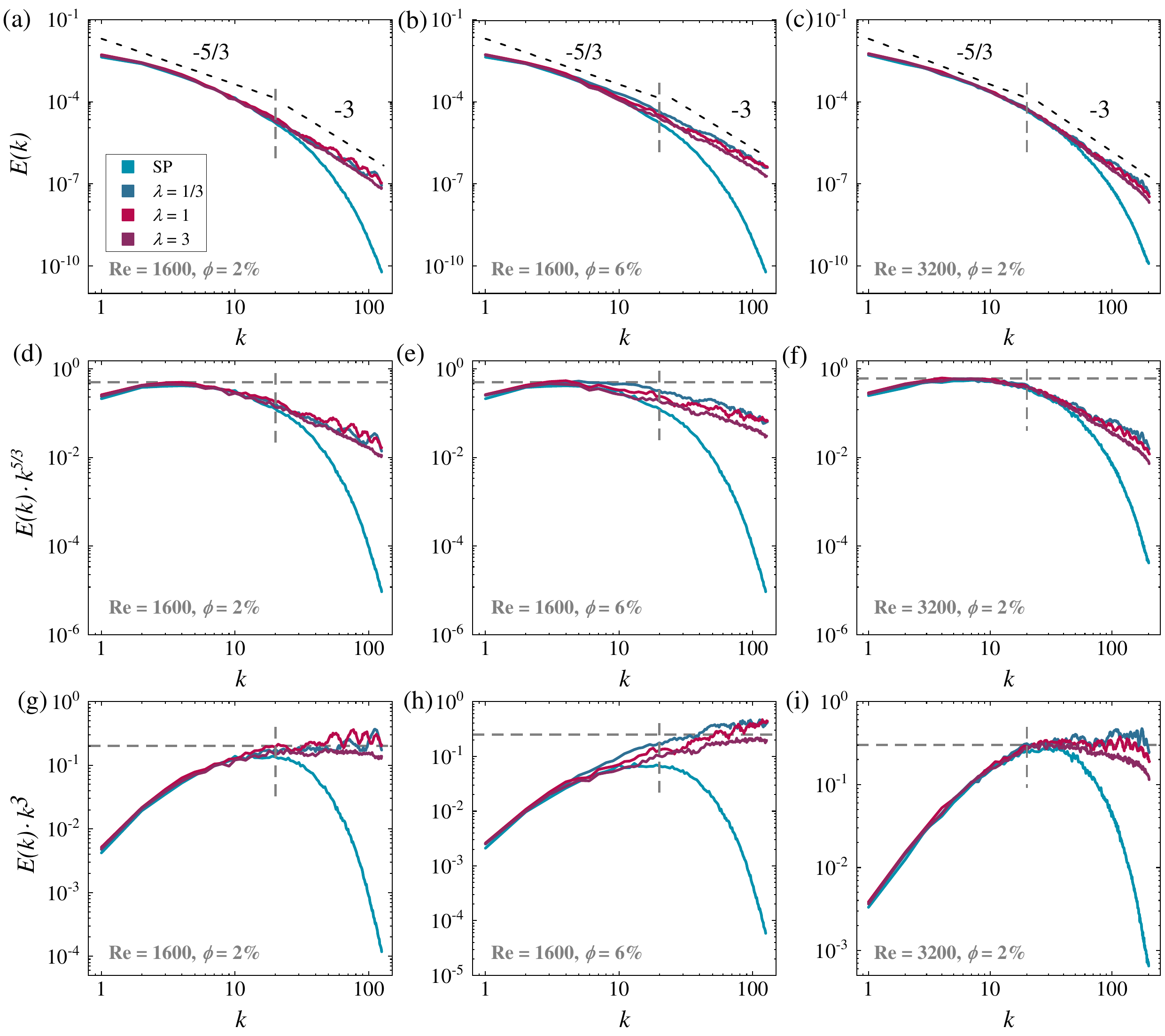}
	\vspace{-5 mm}
	\caption{Turbulent energy spectra (a-c) and the compensated plots ((d-f) are compensated with $k^{-5/3}$, (g-i) are compensated with $k^{-3}$). The two dashed lines shown in the first row are for the guidance of scaling behaviours. The vertical dashed line in each panel indicates the scale of the equivalent-volume diameter of the particles $d_v$.} 
	\label{fig9} 
\end{figure*}

Besides the stress analysis, we also examine the one-dimensional spectra of turbulent velocity fluctuations of the fluid phase for particle-laden cases as shown in figure~\ref{fig9}. 
For all cases of $Re$ and $\phi$ shown in figures~\ref{fig9}(a-c), one can see that the cases of particle-laden flow coincide with that of single-phase flow at large scales while deviating above it at small scales. 
As indicated by the vertical dashed line in each panel, the suspended particles mainly enhance the energy fluctuation at scales smaller than $d_v$. 
In contrast, the particle aspect ratios only mildly affect the spectrum.
For comparison purposes, we show the compensated plot of the power spectrum with $k^{-5/3}$ (figures~\ref{fig9}(d-f)) and $k^{-3}$ (figures~\ref{fig9}(g-i)). 
Both the spectrum of single-phase flow and particle-laden flow follow the well-known scaling $k^{-5/3}$ in an inertial range as proposed by Kolmogorov \citep{pope2000turbulent}. 
It should be noted that the emergence of $k^{-5/3}$ scaling holds over less than one decade, which is mainly due to the moderate turbulence intensity in the present work.
At scales smaller than $d_v$, however, the cases of particle-laden flow show good agreements with the scaling $k^{-3}$. 
Such steeper scaling has been robustly observed in bubbly flow, through experiments \citep{mercado2010bubble,riboux2010experimental,mendez2013power, Prakash2016, Almeras2017, Shao2020, dung2022emergence} and numerics \citep{pandey2020liquid,innocenti2021direct,hidman2022assessing, Perlekar2022}, at scales approximately below the bubble diameter and above the dissipation length scale \citep{hidman2022assessing}. The emergence of $k^{-3}$ in bubbly flow could be attributed to the wake behind bubbles\citep{Almeras2017,risso2018agitation,wang2019self,pandey2020liquid} - also referred to as bubble-induced agitations or pseudo-turbulence - where the velocity fluctuation produced by bubbles is directly dissipated by viscosity \citep{lance1991turbulence,risso2018agitation}. Indeed, bubbles need to be finite-size (i.e. large enough) to generate wakes and thereby the emergence of $k^{-3}$ scaling, typically with a bubble Reynolds number $10\leq Re_{bub}\leq1000$ \citep{risso2018agitation,pandey2020liquid}. In contrast, the $k^{-3}$ scaling could not be observed in simulations of point-like particles since the wake is absent \citep{mazzitelli2009evolution}.

\begin{figure*}
	\centering
	\includegraphics[width=1\linewidth]{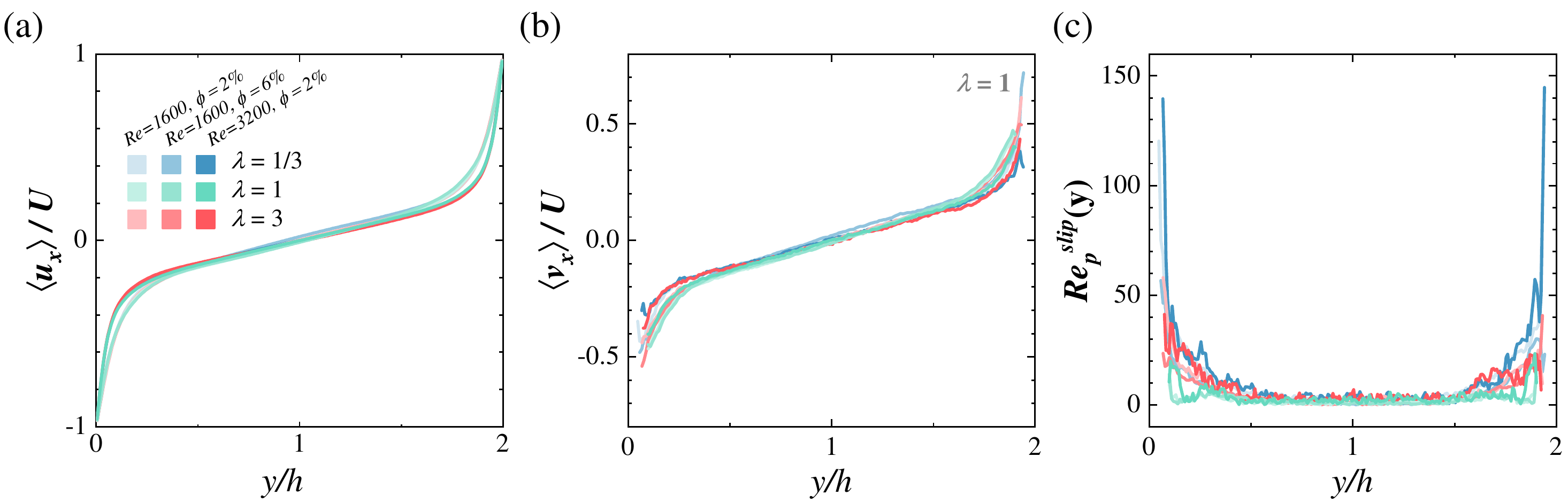}
	\vspace{-5 mm}
	\caption{Profiles of the normalized stream-wise velocity of (a) fluid phase $\langle u_x \rangle$ and (b) particle phase $\langle v_{x} \rangle$. (c) Profile of the particle Reynolds number based on the horizontally averaged slip velocity.} 
	\label{fig10} 
\end{figure*}

However, this seems not to be the situation here for the particle-laden flow even though the particles are finite-size. Unlike the cases of rising bubbles where the relatively large slip velocity is the major contribution to the pseudo-turbulence \citep{risso2018agitation}, the particles used in this work could only yield in small slip velocity due to their neutral buoyancy. Here, we measure the particle Reynolds number based on the mean slip velocity, $Re^{slip}_p = (\langle u_x \rangle - \langle v_{x} \rangle) d_v/\nu$. As shown in Table \ref{table1}, the $Re_p^{slip}$ over the whole domain for each case is of the order $O(1)$ to $O(10)$, which is of marginal value to observe the scaling of $k^{-3}$ for the bubbly flow \citep{risso2018agitation,pandey2020liquid}. However, the particle Reynolds number could be sensitive to the position as the flow is inhomogeneous.
To check this, we look into the profiles of the averaged stream-wise velocity of both phases which are shown in figure~\ref{fig10}. As it can be seen in figure~\ref{fig10}(a,b), the velocity of the particles differs mainly in the near-wall region from the main flow of the fluid. Correspondingly, we measure the profiles of the particle Reynolds number based on the horizontally averaged slip velocity, $Re^{slip}_p(y)$, as shown in figure~\ref{fig10}c.
It can be seen that when close to the walls the local particles Reynolds number is high due to the large slip velocity in this region. Whereas in the bulk region, the $Re^{slip}_p$ is rather small and could be due to their neutral buoyancy. 
This being said the emergence of $k^{-3}$ scaling in the particle-laden flow here could be originated from a different mechanism than that of bubbly flow, which is beyond the scope of the current work.
More studies are needed to further investigate the origin of the `$-3$' scaling observed in particle-laden flow. 

\subsection{Particle statistics: distributions and orientations}\label{sec:third}

\begin{figure*}
	\centering
	\includegraphics[width=1\linewidth]{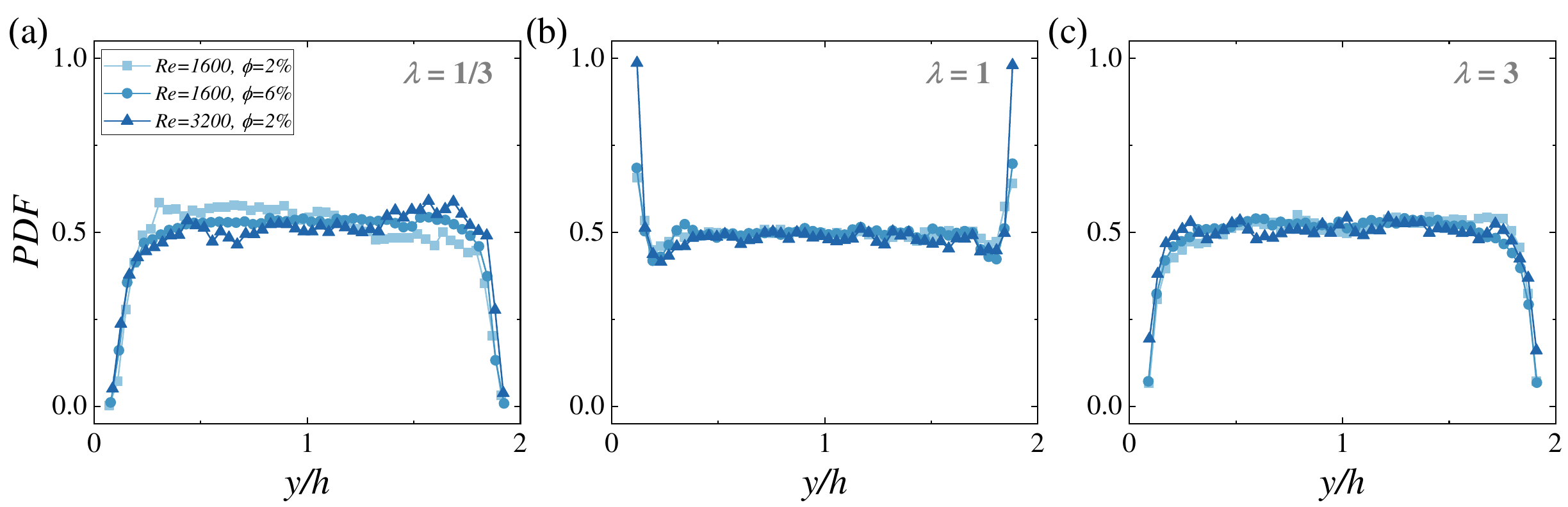}
	\vspace{-5 mm}
	\caption{Particle distributions in the wall-normal direction. Panels from left to right: $\lambda=1/3$, $\lambda=1$ and $\lambda=3$. 
	} 
	\label{fig11} 
\end{figure*}

The particle statistics, as the origins of turbulence modulation, are crucial for understanding the underlying physics. In the section on stress analysis, we find that the spherical particles cause stronger particle-induced stress than the non-spherical ones in the near-wall region. To understand this, we first examine how the particle distributions are affected by the aspect ratio. 
To give a straightforward comparison to previous experimental results \citep{wang2022finite}, we note that here the particle distribution is presented by showing its probability density function (PDF) with respect to its coordinates in wall-normal direction (i.e., $PDF(y)\sim y$), instead of evaluating the local volume fraction by computing the percentage of the grids inside the particle boundary (see e.g. \cite{wang2018transport}).
It can be seen that, as shown in figure~\ref{fig11}, the aspect ratio of particles determines the particle distribution regardless of $Re$ and $\phi$. The spherical particles (figure~~\ref{fig11}b) strongly cluster just near the walls, followed by ebbs next to the peaks. On the contrary, the probability of non-spherical particles appearing in the near-wall region is almost zero (figures~~\ref{fig11}(a, c)), while some above-average peaks occur in the bulk region, suggesting mild but noticeable bulk-cluster effects. Nevertheless, the particles are generally uniformly distributed in most of the bulk. This could be attributed to the gentle turbulence and the less intensive mixing caused by the flows in this region, which would result in a weaker finite-size effect for the particles in the bulk region and thereby weakened clustering.
These clustering effects agree with our experimental findings in \cite{wang2022finite}, and similar results of particle preferential clustering have also been reported in other flows \citep{picano2015turbulent,ardekani2017drag,ardekani2019turbulence}. On the other hand, it is found that, in figure~\ref{fig11}, $Re$ has minor effects on the particle distributions. This could be due to the relatively strong finite-size effects of particles, which allows the particles to escape from the flow structures. 
It should be noted that the particle distribution could depend on $Re$ if the turbulence intensity changes, for example, \cite{wang2018transport} found that the particles could be trapped inside the vortex and show different profiles of distributions at smaller $Re$.

\begin{figure*}
	\centering
	\includegraphics[width=1\linewidth]{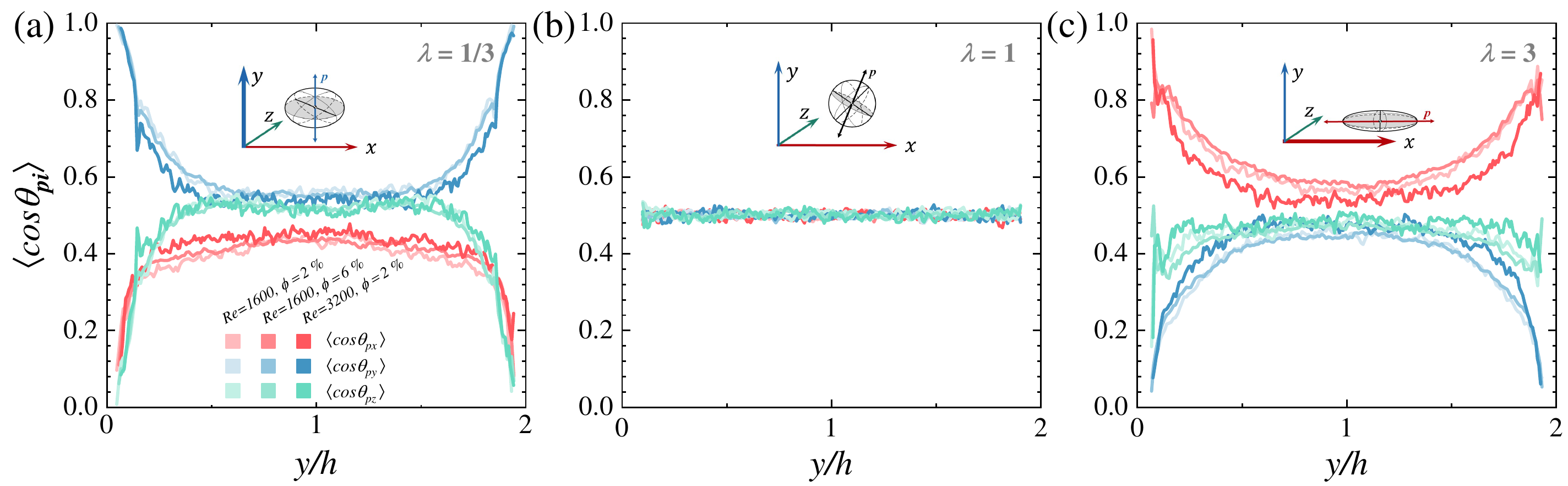}
	\vspace{-5 mm}
	\caption{Profiles of particle orientations in the wall-normal direction. $\theta_{pi}$ is the angle between the symmetric axis of particles $\boldsymbol{p}$ and the coordinate axis $\boldsymbol{i}$. The sketch in each panel are shown to illustrate the preferential alignment of particles in the near-wall region.} 
	\label{fig12} 
\end{figure*}

\begin{figure*}
	\centering
	\includegraphics[width=1\linewidth]{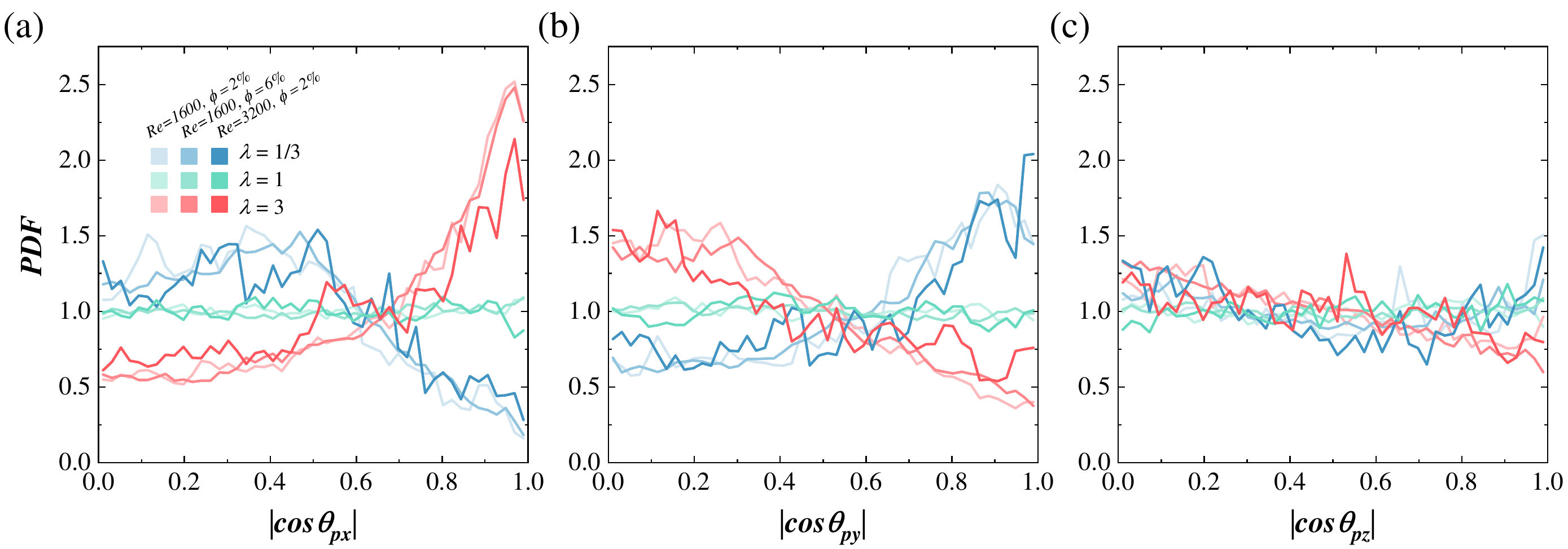}
	\vspace{-5 mm}
	\caption{Probability density functions (PDFs) of the particle orientation in the near-wall region ($y^+ \leq d_v^+$): (a)$|cos\theta_{px}|$, (b) $|cos\theta_{py}|$, (c) $|cos\theta_{pz}|$.} 
	\label{fig13} 
\end{figure*}

In addition to the particle distributions, its orientations (figure~\ref{fig12}) are analyzed. As expected, see figure~\ref{fig12}b, the orientation profiles of spherical particles show no preferential alignment to any coordinate axis since the symmetric axis of spherical particles can be in an arbitrary direction. However, the situation is different for non-spherical particles, particularly in the near-wall regions. For oblate cases (figure~\ref{fig12}a), the symmetric axes of particles show a stronger alignment to the $y$-direction (i.e. wall-normal direction) as the particle location approaches the walls. While for prolate cases (figure~\ref{fig12}c), the particles align their symmetric axes to the $x$-direction (i.e. stream-wise direction) when close to the walls. As in the bulk region, the particles with different aspect ratios still show preferential alignment with the axis as in the near-wall region, i.e., the symmetric axes of oblate (prolate) particles align with the $y$-direction ($x$-direction), while the angles between the symmetric axes of particles and the coordinate axis are almost independent of the particle position. 
In particular, we zoom in on the near-wall region and present the particle preferential orientation in figure~\ref{fig13}. The results show that the oblate (prolate) particles in the near-wall region tend to orientate their symmetric axis to the wall-normal (streamwise) direction, while in the spanwise direction (figure~\ref{fig13}c), particles of all explored cases show noteless preferential orientations due to the unbounded periodic flow in $z-$direction.

\section{Conclusion}
We report a numerical study on suspensions of neutrally buoyant finite-size particles in turbulent plane-Couette flow by employing a fully particle-resolved method. The shape of particles varies from oblate to prolate by changing their aspect ratio $\lambda$. The effects of particle aspect ratio on turbulence modulation (in global arguments and local flows) and particle statistics are studied, which agree well with previous findings in experiments \citep{wang2022finite}. 

The turbulence modulation induced by suspended particles is investigated by changing the aspect ratio $\lambda$, the volume fraction $\phi$ of the particles and the bulk Reynolds number $Re$. For given $Re$ and $\phi$, it is found that the spherical particles could cause the maximum drag increase compared to non-spherical ones. We examine the field of fluid velocity and the dissipation rate field and found that, at given $Re$ and $\phi$, the spherical particles impact the boundary layer stronger than the non-spherical ones. Through stress analysis, we found that the particle-induced stress $\tau_p$ accounts for a great contribution to the total wall stress, particularly in the near-wall region. In addition, the magnitude of $\tau_p$ is larger for the cases of spherical particles than that of non-spherical ones, which, as shown by the particle statistics, could be attributed to the preferential cluster of spherical particles in the near-wall regions. The contribution of each term to the total stress in the stress balance is discussed by integrating them within the whole domain. As a single-phase flow does, the viscous stress and the turbulent stress of the fluid phase still account for the major contributions to total stress in particle-laden flows. However, for the latter cases, the global drag increases could be attributed to the presence of particles, which give rise to the particle-phase turbulent stress and the particle-induced stress. As $Re$ increases, the stress of the fluid phase, particularly the turbulent stress $\tau_{Tf}$, becomes dominant, which results in minor contributions coming from the particle phase and thereby the smaller modulations in global drag.

The turbulent energy spectrum of particle-laden flow is discussed and the results are found to be interesting. At the larger scale, the spectrum of the cases of particle-laden collapse with that of single-phase flow and both follow the scaling of $k^{-5/3}$, suggesting that the flow at large scales is less affected by the presence of particles. While at scales smaller than the equivalent-volume diameter of the particles ($d_v$), the spectrum of particle-laden flow is found to decrease at slower rates than that of single-phase flow. Through compensated plots, we found that the spectrum of particle-laden flow shows good agreement with the scaling of $k^{-3}$ below $d_v$, which is reminiscent of the observation in bubbly flow. 
However, by checking the particle Reynolds number $Re_p^{slip}$, it is found that the $Re_p^{slip}$ ( the order of $O(1)$ to $O(10)$) is too small for the particles to affect the surrounding flows as the bubble does. Therefore, future studies are  required to understand the similarity in the small scale of the turbulent energy spectrum found between the particle-laden flow and bubbly flow, for instance, from the view of particle/bubble dynamics.

Additionally, particle statistics in different regions are found to be affected by the aspect ratios of particles. Spherical and non-spherical particles are found to preferentially cluster in the near-wall and bulk regions, respectively, which are in line with findings in previous experimental results \citep{wang2022finite}. The different cluster effects, which are associated with the intensity of particle collision rate, could account for the magnitudes of particle-induced stress and the resulting global drag. On the other side, spherical particles, as expected, show no preferential alignment to any direction due to their perfect symmetry. Whereas, oblate and prolate particles are found to preferentially align with the wall-normal and stream-wise direction due to their anisotropy. 

In collaboration with our previous experimental findings, this numerical study provides deeper sights into the information of both fluid and particle phases, allowing one to understand how the turbulence is modulated by the suspended particles. 
Limited by computation consumption, the present work mainly focuses on the role of particle aspect ratio, while the particle volume fractions are kept at relatively small values. It is expected in future work to observe greater turbulence modulation in the dense regime of particle suspensions. 
\\

\noindent {\bf Acknowledgements:}
We thank Enrico Calzavarini for the help with the numerical simulations \& the insightful discussions, and thank Jinghong Su, Dongpu Wang, Lei Yi and Ning Zhu for the helpful discussions. \\

\noindent {\bf Funding:}
This work was supported by the National Natural Science Foundation of China under grant no. 11988102, and the Tencent Foundation through the XPLORER PRIZE.\\

\noindent {\bf Declaration of interests:} The authors report no conflict of interest.\\

\appendix 
\section{Fluctuation of fluid and particle velocity }\label{sec:appendix}
\begin{figure*}
	\centering
	\includegraphics[width=1\linewidth]{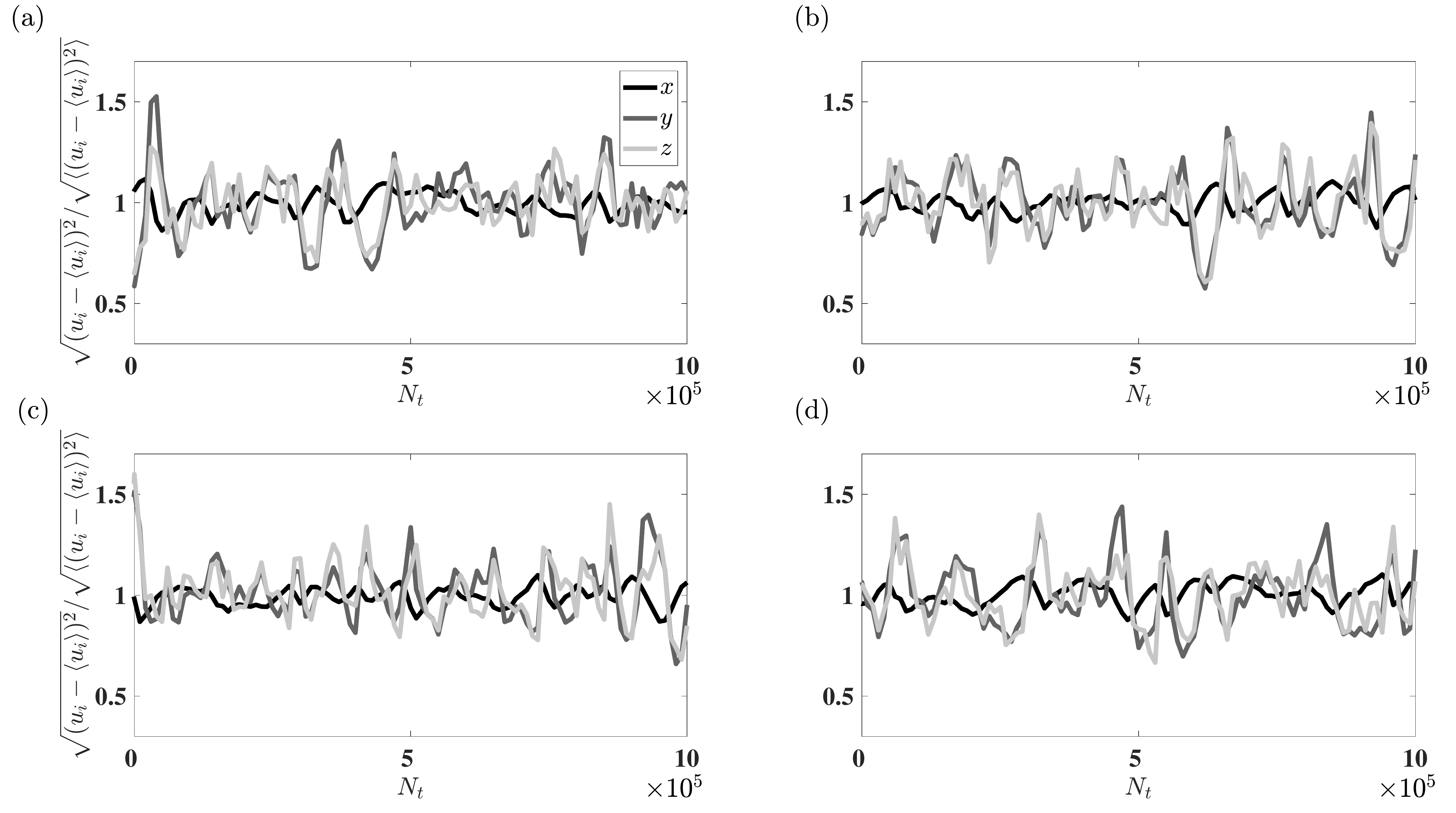}
	\vspace{-3 mm}
	\caption{Fluid velocity fluctuation verse time at $Re = 1600$: (a) Single-phase case, (b) $\lambda=1/3$, (c) $\lambda=1$, (d) $\lambda=3$. For particle-laden cases, $\phi=2\%$. The horizontal axis is denoted by the numerical time steps $N_t$, and the vertical axis is the normalized fluid velocity fluctuations.} 
	\label{fig14} 
\end{figure*}

\begin{figure*}
	\centering
	\includegraphics[width=1\linewidth]{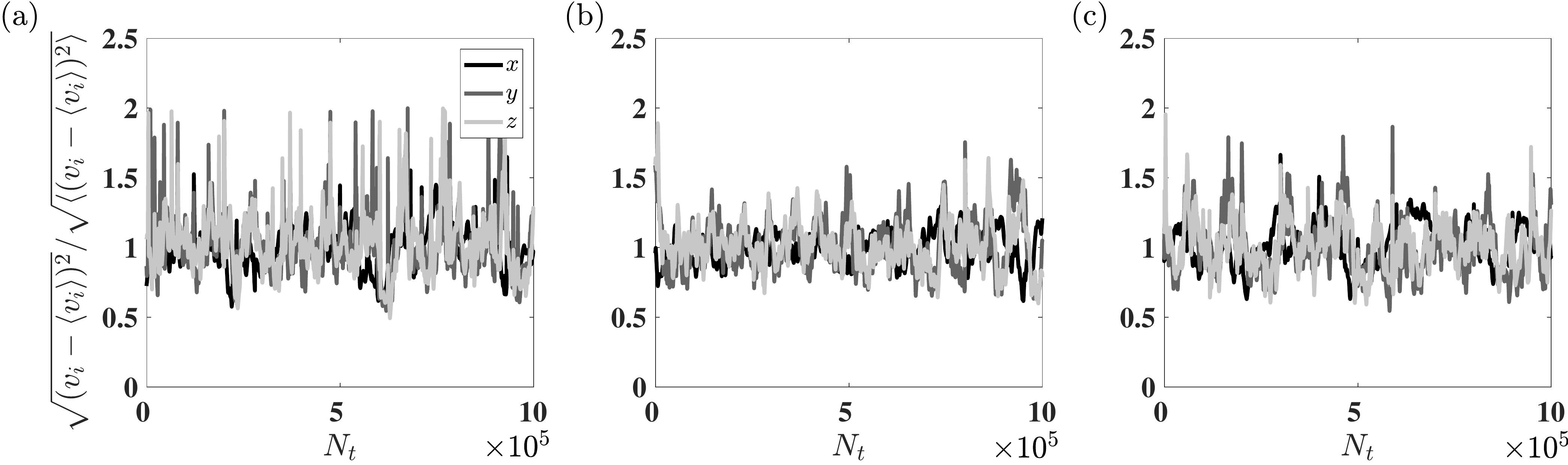}
	\vspace{-5 mm}
	\caption{Particle velocity fluctuation verse time at $Re = 1600$ and $\phi=2\%$: (a) $\lambda=1/3$, (b) $\lambda=1$, (c) $\lambda=3$. The horizontal axis is denoted by the numerical time steps $N_t$, and the vertical axis is the normalized particle velocity fluctuations.} 
	\label{fig15} 
\end{figure*}

{
    Since the fluid and particle velocity fluctuations contribute to the stress decomposition in equation~\ref{stressbudget}), it is obligatory to check how they converged in time. The results are shown in figure~\ref{fig14} and figure~\ref{fig15}. Only the cases at $Re=1600$ and $\phi=2\%$ are shown here. For all cases with/without particles, the fluid velocity fluctuates within reasonable ranges. As expected the maximum fluctuation occurs in the streamwise direction since the flow is driven by the horizontally moving walls. The fluctuations in wall-normal and spanwise directions are rather small. The fluctuations of particle velocity (figure~\ref{fig15}) are similar to those of the fluid phase but with stronger amplitudes. Additionally, some burst events in the particle velocity fluctuation occur, which may be due to stronger particle collisions. To conclude, these figures show that the velocity fluctuations of both phases converged in time.
}

\end{document}